\documentclass[12pt]{article}
     
\usepackage{latexsym}
\usepackage{amssymb}
\usepackage{amsmath}
\usepackage{dsfont} 
\usepackage{graphicx}
\usepackage{bbm}
\usepackage{color}
\input{colordvi} 

\newcommand{\comment}[1]{}

\newcommand{\ran}{\mathop{\mathrm{ran}}}
\newcommand{\id}{\mathbbm{1}}
\newtheorem{thm}{Theorem}
\newtheorem{lem}[thm]{Lemma}
\newtheorem{prop}[thm]{Proposition}

\newtheorem{rem}{Remark}

\newtheorem{ass}{Assumption}

\newcommand{\ket}[1]{\left| #1 \right\rangle}
\newcommand{\bra}[1]{\left\langle #1 \right|}
\newcommand{\braket}[2]{\langle #1 | #2 \rangle}
\newcommand{\pder}[1]{\frac{\partial}{\partial #1 }}
\renewcommand{\d}{\mathrm{d}}

\definecolor{light}{gray}{.75}

\title{Adiabatic theorem for a class of quantum stochastic equations}
\author{M. Fraas
\\
\small{Theoretische Physik, ETH Z\"{u}rich, 8093 Z\"{u}rich, Switzerland}
 }

\begin{document}

\maketitle
\begin{abstract}
We derive an adiabatic theory for a stochastic differential equation,
$$ \varepsilon\, \d X(s) = L_1(s) X(s)\, \d s + \sqrt{\varepsilon} L_2(s) X(s) \, \d B_s, $$
under a condition that instantaneous stationary states of $L_1(s)$ are also stationary states of  $L_2(s)$. We use our results to derive the full statistics of tunneling for a driven stochastic Schr\"{o}dinger equation describing a dephasing process.
\end{abstract}

We study solutions of a stochastic differential equation
\begin{equation}
\label{eq:1}
\varepsilon\, \d X(s) = L_1(s) X(s)\, \d s + \sqrt{\varepsilon} L_2(s) X(s) \, \d B_s, \quad s \in (0,\,1)
\end{equation}
where $L_1,\,L_2$ are bounded operators on a Hilbert space $\mathcal{H}$ and $B_s$ is a Brownian motion. The equation is expressed in the slow time $s = \varepsilon t$. The scaling of the second term reflects the Brownian scaling, $\varepsilon^{-1/2} B_{\varepsilon^{-1} s}$ is a Brownian motion in the slow time for any $\varepsilon > 0$. The adiabatic theory studies solutions of the equation in the limit $\varepsilon \to 0$.

A particular example of Eq.~(\ref{eq:1}) that motivates our study is a slowly driven stochastic Schr\"{o}dinger equation, a classical member of the family of quantum stochastic equations derived by Hudson and Parthasarathy \cite{Hudson}. In their full extent quantum stochastic equations describe a system linearly coupled to a bosonic free field. When the coupling is through the position or momentum operator only the equations are equivalent to classical \^{I}to equations with the Brownian motion representing the bath. Conditioning the dynamics on a continuous measurement on the free field gives non-linear quantum filtering equations derived by Belavkin \cite{Belavkin}. These equations (and their time-discrete counterparts) provide basic framework for quantum closed loop feedback and control \cite{Wiseman, Gough}.  The goal of our line of research is to develop a feedback theory for the adiabatic quantum control. In particular we plan to develop an adiabatic theory for quantum filtering equations. The adiabatic theory for the unconditioned stochastic Schr\"{o}dinger equation, derived here, is the first step in this direction. 

Equation~(\ref{eq:1}) has been widely studied in the deterministic case, $L_2(s) \equiv 0$, see \cite{Joye, AFGG, S} and references therein. The main feature of the adiabatic theory is that solutions of Eq.~(\ref{eq:1}) can be described algebraically as follows.
\begin{enumerate}
\item[(i)] The evolution generated by the equation leaves the kernel of $L_1$ invariant to the leading order in $\varepsilon$;
\item[(ii)] There is an asymptotic expansion that describes the motion inside the kernel and the  tunneling out of the kernel. 
\end{enumerate}
Leaving aside exact assumptions, it was understood by Avron and Elgart \cite{AE99} that (i) holds provided one can define the projection on the kernel in a continuous manner. On the other hand (as was long known), the expansion (ii) exists provided $0$ is an eigenvalue isolated from the rest of the spectra (so called gap condition). We will study only the case when the projection can be defined continuously irrespectively of the Brownian path. Hence a generalization of (i) might not be surprising. However, we will also derive an expansion (ii), which is somehow surprising because the gap condition cannot hold for all realizations of the Brownian motion. 

The most restrictive condition of our theory is a requirement that $\ker L_1(s) \subset \ker L_2(s)$ for each instant of time. Under this assumption we derive below an asymptotic expansion for the solution of a form
$$
X(s) = x_0(s) + \sqrt{\varepsilon} y_1(s) + \varepsilon x_1(s) + \dots + \varepsilon^{N-\frac{1}{2}} y_N(s) + \varepsilon^N x_N(s) + O(\varepsilon^{N+1/2}).
$$
The standard integer power terms ($x$'s) are deterministic and given by the adiabatic expansion in the absence of the stochastic term, $L_2 = 0$. The novel half integer terms are stochastic and describe propagation of an instantaneous error to the future. They are expressed as backward \^{I}to integrals arising from a Duhamel formula.

\comment{
In the deterministic case, the motivation to study Eq.~(\ref{eq:1}) on a Banach space came from an adiabatic Lindblad equation, \cite{Lidar}, describing a slowly driven open quantum system. Our motivation to study the stochastic version of Eq.~(\ref{eq:1}) is the stochastic Schr\"{o}dinger equation, which is an unraveling of the Lindblad equation. For this application it would be sufficient to study the adiabatic theory on a Hilbert space. Nevertheless we chose to formulate our main results on the Banach space because this generality comes with zero additional costs. The application to a driven stochastic Schr\"{o}dinger equation is described in details in Section~\ref{Schrodinger}.
}

In the case of stochastic Schr\"{o}dinger equation with a simple ground state\footnote{Or any simple isolated eigenvalue.} the stochastic term of order $\sqrt{\varepsilon}$ is orthogonal to the ground state and describes the tunneling out of the ground state. We derive a formula for this tunneling and describe its full statistics. This extends the work \cite{AFGG} where a formula for the mean tunneling was derived by studying a slowly driven Lindblad equation \cite{Lidar}. These two equations are closely connected, the latter is obtained from the stochastic Schr\"{o}dinger equation by averaging over the randomness.

The article is organized as follows. In the remaining part of the introduction we introduce our notation and discuss basics of the stochastic calculus necessary to follow our exposition. In Section~\ref{calculus} we describe the stochastic calculus in more details, in particular we describe the two sided stochastic calculus of Pardoux and Protter \cite{Pardoux}. We also state there several technical propositions regarding the stochastic integration. The reader not interested in proofs might safely skip the section. Section~\ref{basic} gives our assumptions and basic results.  In the following Section~\ref{schrodinger} we apply these results to a stochastic Schr\"{o}dinger equation describing dephasing and derive the full statistics of tunneling in the leading order. The last section contains the full adiabatic expansion and its proof.

\begin{rem}
In view of the application we had in mind we chose to describe the theory on a Hilbert space rather then on a Banach space. Extension to a finite dimensional Banach space is straightforward. Infinite dimensional Banach spaces introduce several technical complications (starting with the very existence of the \^{I}to calculus) and we do not know what are natural assumptions on the geometry of the Banach space for the extension of our results.

We comment on various complications with the Banach space theory throughout the article.
\end{rem}

We denote the scalar product on $\mathcal{H}$ by $(\cdot,\cdot)$ and the norm by $|| \cdot ||$. We suppress randomness from our notation and ``$=,\,\leq,\,\dots$" between random variables holds with probability $1$. $\mathbb{E}[\cdot]$ stands for the expectation value with respect to the Brownian motion, and $||\cdot||_\infty$ is the corresponding $L^\infty$ norm. In particular for a random variable $X \in \mathcal{H}, ||X|| \leq ||X||_\infty$. $O(\varepsilon^n)$ is a random variable for which $\varepsilon^{-n} ||O(\varepsilon^n)||, \varepsilon \in (0,1) $ is a family of random variables with uniformly bounded moments. 

 We make extensive use of \^{I}to calculus and recall that for non-anticipatory functions $f,\,g$ it holds that
$$
\d (fg) = \d f g + f \d g + \d f \d g,\quad d(f \circ g) = (f' \circ g) \d g + \frac{1}{2} (f'' \circ g) (\d g)^2,
$$
where $\d f \d g$ should be interpreted according to the rules $(\d s)^2 = \d s \d B_s = 0$, $(\d B_s)^2 = \d s$. We also use the backward \^{I}to calculus which comes with a  similar set of rules  given in the following section. In a nutshell backward \^{I}to calculus integrates functions of the future, while \^{I}to forward calculus integrates functions of the past. 

 To illuminate the difference between the forward/backward integrals we consider the two-parameter stochastic propagator \cite{Skorokhod}, $U_{\varepsilon}(s,\,s')$, associated to Equation~(\ref{eq:1}). This is a random variable that depends on the Brownian increments in the interval $(s',\,s)$. As a function of $s$, for a fixed $s'$, the propagator satisfies a forward \^{I}to equation,
 \begin{equation}
 \label{forward}
 U_\varepsilon(s,\,s') = \id + \int_{s'}^s L_1(t) U(t,\,s') \d t + \int_{s'}^s L_2(t) U(t,\,s') \d B_t.
 \end{equation}
 On the other hand as a function of $s'$ it satisfies a backward \^{I}to equation
 \begin{equation}
 \label{backward}
 U_\varepsilon(s,\,s') = \id + \int_{s'}^s  U(s,\,t) L_1(t) \d t + \int_{s'}^s  U(s,\,t) L_2(t) \d B_t.
 \end{equation}
 We will not stress the difference between the backward and the forward integration in our notation. If the integrand refers to the past (it is non-anticipatory) it is a forward integral, if the integrand refers  to the future it is a backward integral. In fact we use a shorthand differential notation,
 \begin{align*}
 \varepsilon\, \d_s U_\varepsilon(s,\,s') &= \d L(s) U_\varepsilon(s,\,s'),  &U_\varepsilon(s',\,s')  = \id ,\\ \quad \varepsilon\, \d_{s'} U_\varepsilon(s,\,s') &= -U_\varepsilon(s,\,s') \d L(s'),  &U_\varepsilon(s,\,s)  = \id, 
 \end{align*}
 as an equivalent of Eq.~(\ref{forward}) respectively Eq.~(\ref{backward}), where $\d L(s) = L_1(s) \d s + \sqrt{\varepsilon} L_2(s) \d B_s$.
 
 We end this short exposition with two standard relations that hold for both forward and backward integration,
 \begin{equation}
 \label{itometry}
 \mathbb{E}[ \int_0^1 X_t \d B_t ] = 0, \quad \mathbb{E}[ || \int_0^1 X_t \d B_t||^2] = \int_0^1 \mathbb{E}[||X_t||^2] \d t,
 \end{equation}
 the latter equality can be understood using a formal relation $\mathbb{E}[\d B_t \d B_s] = \delta(t-s) \d t$.

\section{A two-sided stochastic calculus}
\label{calculus}

We are neither competent nor it is our purpose to explain the stochastic calculus in this preliminary section. Several textbook and monographs are devoted to this topic, the author used a book of McKean \cite{McKean} and the relevant chapters in a book of Simon \cite{Simon}. We do not even aim to explain the two sided integral constructed by Pardoux and Protter \cite{Pardoux}, the reader should consult their article for details and proofs. We merely repeat what is relevant for our exposition and we gather several lemmas that we shall need for the proofs.

We consider a one dimensional Brownian motion $B_t,\, 0 \leq t \leq 1$ and the associated sigma algebra $\sigma(B_s,\,0 \leq s \leq t)$. For a continuous process $X_t$ adapted to the sigma algebra the forward \^{I}to integral of $X_t$ is defined as
$$
\int_0^s X_t \d B_t = \lim_{N \to \infty} \sum_{k=1}^{2^N} X_{\frac{k-1}{2^N}s}(B_{\frac{k}{2^N}s} - B_{\frac{k-1}{2^N}s}).
$$
It is an important part of the definition that the increment points to the future and hence $B_{\frac{k}{2^N}s} - B_{\frac{k-1}{2^N}s}$ and $X_{\frac{k-1}{2^N}s}$ are independent random variables. A consequence of this choice is that the integral, as a function of $s$, is a martingale and two basic formulas, cf. (\ref{itometry}),
$$
\mathbb{E}[\int_0^s X_t \d B_t] = 0, \quad \mathbb{E}[\left( \int_0^s X_t \d B_t \right)^2] = \int_0^s \mathbb{E}[X_t^2] \d t,
$$
hold true.

Backward \^{I}to integral is defined in an analogous manner. We consider a process $Y_t$ adapted to a sigma algebra $\sigma(B_s - B_1,\, t \leq s \leq 1)$ and we define the integral of $Y_t$ by
$$
\int_s^1 Y_t \d B_t = \lim_{N \to \infty} \sum_{k=1}^{2^N} Y_{s + (1-s) \frac{k}{2^N}} (B_{s + (1-s) \frac{k}{2^N}} - B_{s + (1-s) \frac{k-1}{2^N}}).
$$
Note that here the increments point to the future in order to ensure the independence with the integrand. The backward \^{I}to integral is a backward martingale as a function of $s$ and formulas corresponding to (\ref{itometry}) hold true,
$$
\mathbb{E}[\int_s^1 Y_t \d B_t] = 0, \quad \mathbb{E}[\left( \int_s^1 Y_t \d B_t \right)^2] = \int_s^1 \mathbb{E}[Y_t^2] \d t.
$$

We now consider particular processes $X_t,\,Y_t$ that arise as solutions of forward/backward stochastic differential equations,
\begin{align*}
X_t &= X_0 + \int_0^s b(X_t) \d t + \int_0^s \sigma(X_t) \d B_t ,\\
Y_t &= Y_0 + \int_s^1 c(Y_t) \d t + \int_s^1 \gamma(Y_t) \d B_t, \\ 
\end{align*} 
for some continuous functions $b,\,c,\,\sigma,\,\gamma$. The integral in the first equation being the forward \^{I}to integral, while the integral in the second equation being the backward \^{I}to integral. Correspondingly the first equation has a unique solution $X_t$ that is a non-anticipatory martingale and the second equation has a unique solution $Y_t$ that is a backward martingale adapted to the associated sigma algebra.

We also use a differential form of these equations
$$
\d X_t = b(X_t) \d t + \sigma(X_t) \d B_t, \quad \d Y_t = -c(Y_t) \d t - \gamma(Y_t) \d B_t.
$$
Although the notation makes no distinction between the forward and the backward case one should keep in mind that these are distinct differentials.

A stochastic integral for joint functions of $X_t,\,Y_t$ was constructed in \cite{Pardoux}. Let $f(t,\,X_t,\,Y_t)$ be a continuous function of its arguments, then an integral
$$
\int_{s'}^s f(t,X_t,Y_t) \d B_t 
$$
can be defined in such a way that if $f$ is independent of $Y_t$ (resp. $X_t$) then the integral coincides with the forward (resp. backward) \^{I}to integral.  Furthermore  the integral satisfies the following chain rule,
\begin{multline}
\label{chainrule}
f(s,\,X_s,\,Y_s) = f(s',\,X_{s'},\,Y_{s'}) + \int_{s'}^s \partial_t f(t,X_t,Y_t) \d t \\ + \int_{s'}^s \partial_{X} f(t,X_t,Y_t) \d X_t  + \int_{s'}^s \partial_{Y} f(t,X_t,Y_t) \d Y_t \\ + \frac{1}{2} \int_s^{s'} \partial_{XX} f(t,X_t,Y_t) (\d X_t)^2 - \frac{1}{2} \int_s^{s'} \partial_{YY} f(t,X_t,Y_t) (\d Y_t)^2,
\end{multline}
where $(d X_t)^2 = \sigma^2(X_t) \d t$ is interpreted according to the \^{I}to rules $(\d t)^2 = \d t \d B_t = 0$, $(\d B_t)^2 = \d t$.

In the following paragraphs we employ the formula~(\ref{chainrule}) in a case with no second order derivatives to operator valued processes $X_t,\,Y_t$. Due to the simplicity of that case the operator valued extension is clear. The operator valued version is discussed in more details in \cite{Pardoux}.

To demonstrate the power of the calculus we show that equations (\ref{forward}, \ref{backward}) define the same propagator (the value of $\varepsilon$ is not important for the following considerations and we skip the index) and that this propagator satisfies the semigroup property
$$
U(s,0) = U(s,s') U(s',0).
$$
To this end we fix a time $t$ and let $X_s = U(s,0)$ be a solution of Eq.~(\ref{forward}) and $Y_s= \tilde{U}(t,s)$ be a solution of Eq.~(\ref{backward}). Then the above chain rule implies that for any $t \leq s \leq s'$ we have
$$
\tilde{U}(t,s) U(s,0) = \tilde{U}(t,s') U(s',0).
$$
By choosing $s = t$ and $s'=0$ we get the sought equivalence $\tilde{U}(t,0) = U(t,0)$. Upon erasing the tilde in the above equation we then establish the semigroup property.

In the following we will need two specific  results concerning stochastic differential equations. The first is a particular version of the Duhamel formula, the second is a  prior bound on stochastic integrals. We formulate the bound for the forward integral, the corresponding bound holds also for the backward integral.

\begin{lem}[Duhamel formula] 
\label{duhamel}
The solution of the differential Eq.~(\ref{forward}) satisfies a relation
$$
U(s,\,s') = V(s,s') + \int_{s'}^s U(s,t) L_2(t) V(t,s') \d B_t,
$$
where $V(s,s')$ is the solution of a deterministic equation,
$$
\d V(s,s') = L_1(s) V(s,s') \d s, \quad V(s',s') = \id.
$$
\end{lem} 
{\bf Proof:} The proof is again an application of the chain rule (\ref{chainrule}). Pick $X_t = V(t,s'),\, Y_t = U(s,t)$ then for any $s \geq t \geq t' \geq s'$ the chain rule gives
$$
U(s,t) V(t,s') = U(s,t') V(t',s') - \int_{t'}^t U(s,x) L_2(x) V(x,s') \d B_x.
$$
The statement of the lemma then follows by choosing $t = s$ and $t' = s'$. \hfill $\square$

\begin{lem}[Prior estimates of stochastic integrals]
\label{bound}
Let $X_t \in \mathcal{H}$ be a non-anticipatory stochastic process, then the following estimates hold true:
\begin{enumerate}
\item[(a)]
$$
\mathbb{E}[ ||\int_0^1 X(s) \d B_s ||^{2n}] \leq (2n^2 - n)^n \mathbb{E}[\int_0^1 ||X_s||^{2n} \d s], \quad n \geq 1.
$$
\item[(b)]
\begin{equation}
\label{expbound}
Prob( || \int_0^1 X_s \d B_s ||^2 > \gamma) \leq e^{-\gamma \frac{1}{8 ||X||_\infty^2} + \frac{1}{4}}, 
\end{equation}
where $||X||_\infty := \sup_{0 \leq t \leq 1} ||X_t||_\infty$.
\end{enumerate}
\end{lem}
{\bf Proof:}
Denote $\Theta_t = \int_0^t X_s \d B_s$ and consider a real valued stochastic process $\zeta_t = (\Theta_t, \Theta_t)$. The stochastic differentiation of this process gives $\d \zeta_t = ((X_t, \Theta_t) + (\Theta_t, X_t)) \d B_t+ (X_t,X_t) \d t$, which is equivalent to an integral relation
\begin{equation}
\label{label}
\zeta_t - \int_0^t ||X_s||^2 \d s = \int_0^t ((X_s, \Theta_s) + (\Theta_s, X_s)) \d B_s.
\end{equation}

(a) We take the expectation of $d\zeta_t^n = n \zeta_t^{n-1} \d \zeta_t + 1/2n(n-1) \zeta_t^{n-2} \d \zeta_t \d \zeta_t$ to get an integral relation,
$$
\mathbb{E}[\zeta_t^n] = n \int_0^t \mathbb{E}[\zeta^{n-1}_s ||X_s||^2] \d s + \frac{n(n-1)}{2} \int_0^t \mathbb{E}[ \zeta_s^{n-2} ( (\Theta_s, \,X_s) + (X_s,\,\Theta_s))^2] \d s,
$$
between the moments. As a first observation note that all integrands are non-negative functions and hence $\mathbb{E}[\zeta_t^n]$ is a non-decreasing function of $t$. Now we employ the estimate $((X_s, \Theta_s) + (\Theta_s, X_s))^2\leq 4 ||X_s||^2 \zeta_s$ and the H\"{o}lder inequality to get
\begin{align*}
\mathbb{E}[\zeta_t^n] &\leq (n + 2n(n-1)) \int_0^t \mathbb{E}[ \zeta_s^{n-1} || X_s||^2] \d s \\
				 & \leq (2n^2 -n) \left( \int_0^t \mathbb{E}[\zeta_s^n] \d s \right)^\frac{n-1}{n} \left( \int_0^t ||X_s||^{2n} \d s \right)^{\frac{1}{n}} \\
				 &\leq (2n^2 -n) \left( \mathbb{E}[\zeta_t^n] \d s \right)^\frac{n-1}{n} \left( \int_0^t ||X_s||^{2n} \d s \right)^{\frac{1}{n}} .
\end{align*}
In the last inequality we also used $0 \leq t \leq 1$. Solving for the $n$-th moment establishes the first inequality of the lemma.

To prove (b) we will use a well-known prior estimate on stochastic integrals \cite[Chapter 2.3]{McKean}. Let $e_t, 0 \leq t \leq 1$ be a real non-anticipatory function and suppose that $\int_0^1 e_t^2   \d t < \infty$ then for any reals $\beta$ and $\alpha$ the following bound holds true,
\begin{equation}
\label{martin}
Prob\left[ \max_{0 \leq t \leq 1} ( \int_0^t e_s \d B_s  - \frac{\alpha}{2} \int_0^t e_s^2 \d s ) > \beta \right] \leq e^{-\alpha \beta}.
\end{equation}
We note that the bound is an application of Doob's martingale inequality.

Applying the bound to Eq.(\ref{label}) then implies
$$
Prob\left[ \max_{0 \leq t \leq 1} \left( \zeta_t - \int_0^t ||X_s||^2 \d s - \frac{\alpha}{2} \int_0^t ((X_s, \Theta_s) + (\Theta_s, X_s))^2 \d s \right)> \beta \right] \leq e^{-\alpha \beta}.
$$
We claim that for $0 \leq t \leq 1$,
\begin{equation}
\label{stam}
\zeta_t - \int_0^t ||X_s||^2 \d s - \frac{\alpha}{2} \int_0^t ((X_s, \Theta_s) + (\Theta_s, X_s))^2 \d s  \geq \zeta_t - ||X||_\infty^2 - 2 \alpha ||X||_\infty^2 \max_{0 \leq s \leq 1} \zeta_s.
\end{equation}
In particular whenever $\max RHS > \beta$ then also $\max LHS > \beta$ and the probability of an event $\max RHS > \beta$ is smaller then the probability of an event $\max LHS > \beta$. Combining this with the probability bound above we have
\begin{multline*}
Prob\left[ \max_{0 \leq t \leq 1} \zeta_t - ||X||_\infty^2 - 2 \alpha ||X_\infty||^2 \max_{0 \leq t \leq 1} \zeta_t> \beta \right]  \\
= Prob \left[ \max_{0 \leq t \leq 1} \zeta_t > \frac{\beta + ||X||_\infty^2}{1 - 2 \alpha ||X||_\infty^2} \right] \leq e^{- \alpha \beta}.
\end{multline*}
Writing $\gamma = (\beta + ||X||_\infty^2)/(1 - 2 \alpha ||X||_\infty^2)$ and choosing the optimal $\alpha = 1/(4 ||X||_\infty^2)$ we get Bound~(\ref{expbound}).

It remains to prove Eq.~(\ref{stam}). The inequality follows from the inequality $ ||X_s||^2 \leq  ||X||^2_\infty $ and the inequality
$$
((X_s, \Theta_s) + (\Theta_s, X_s))^2\leq 4 ||X||^2_\infty ||\Theta_s||^2   \leq 4 ||X||^2_\infty \max_{0 \leq s \leq 1} \zeta_s.
$$
Note that for $0 \leq t \leq 1$ an integral $\int_0^t$ of a positive constant can be bounded by that constant.
\hfill $\square$ 

An important consequence of the lemma is that
\begin{equation}
\label{Ocalculus}
\int_0^1 O_t(\varepsilon^n) \d B_t = O(\varepsilon^n),
\end{equation}
provided the moments of $||O_t(\varepsilon^n)||$ are uniformly bounded with respect to $t$.

Generalization of Lemma~\ref{bound}. is one of the main technical obstacles of a Banach space version of the theory. For finite dimensional spaces all norms are equivalent and the above bounds hold true up to a dimension dependent constant. On the other hand we do not know if such bounds are available in the infinite dimensional Banach spaces.

The adiabatic expansion, which is the main result of our paper, has a natural formulation in terms of the backward \^{I}to integral.  On the other hand it is often easier -- not principally, just thanks to a larger degree of familiarity -- to perform calculations with a forward \^{I}to integral.  Due to a special structure of integrals that appear in this work we can always convert a backward integral to a forward integral.

\begin{rem}
\label{conversion}
The backward stochastic integrals of the type $\int_0^s U(s,\,s') f(s') \d B_{s'}$, where $f$ is a deterministic function, can be converted into a forward integral thanks to the semigroup relation $U(s,\,s') =U(s,0) U(s',0)^{-1}$. The relation expresses the propagator in the future by a constant (with respect to the integration) times a propagator in the past. We still need to convert the backward to a forward integral.

To see in details how the conversion works we take a second look at the forward \^{I}to integral that we defined by
$$
I_- = \lim_{N \to \infty} \sum_{k=1}^{2^N} X_{\frac{k-1}{2^N}s}(B_{\frac{k}{2^N}s} - B_{\frac{k-1}{2^N}s}).
$$
Alternatively one can define\footnote{This is sometimes referred to as a backward integral, we do not use this name to avoid confusion.}
$$
I_+ = \lim_{N \to \infty} \sum_{k=1}^{2^N} X_{\frac{k}{2^N}s}(B_{\frac{k}{2^N}s} - B_{\frac{k-1}{2^N}s})
$$
and these two definitions are related by a quadratic variation of $X_t$,
$$
I_+ - I_- = \int_0^s \d X_t \d B_t = \int_0^s \sigma(X_t) \d t.
$$
For the integral under consideration this now implies --- we introduce back $\varepsilon$ as this will be useful at a later point in the article ---
\begin{multline}
\label{eq:6}
\int_0^s U_\varepsilon(s,\,s') f(s') \d B_{s'}  - U_\varepsilon (s,0) \int_0^s U_\varepsilon(s',0)^{-1} f(s') \d B_{s'}  \\ =   - \frac{1}{\sqrt{\varepsilon}}U_\varepsilon (s,0) \int_0^s U_\varepsilon(s',0)^{-1} L_2(s') f(s')  \d s',
\end{multline}
where we have used $\varepsilon \d U(s ,0)^{-1}  = U(s ,0)^{-1} (-\d L(s) +L_2^2(s) \d s)$. The second line seems to diverge as $\varepsilon \to 0$, but in fact it is of order $1$, and provided all the inverse operators exist on the range of $L_2$ we have:
\begin{equation}
\label{eq:7}
\int_0^s U_\varepsilon(s,\,s') f(s') \d B_{s'} = U_\varepsilon (s,0) \int_0^s U_\varepsilon(s',0)^{-1} \tilde{f}(s') \d B_{s'} + O(\sqrt{\varepsilon}),
\end{equation}
with $\tilde{f}(s) = [1 + L_2(s) (L_1(s) - L_2^2(s))^{-1} L_2(s)] f(s)$. To see this we use
\begin{multline*}
\int_0^s U_\varepsilon(s',0)^{-1} L_2(s') f(s')  \d s' = \varepsilon \int_0^s \d U_\varepsilon(s',0)^{-1} (-L_1(s') + L_2^2(s'))^{-1} L_2(s') f(s') \\
+ \sqrt{\varepsilon} \int_0^s  U_\varepsilon(s',0)^{-1} L_2(s')  (-L_1(s') + L_2^2(s'))^{-1} L_2(s') f(s') \d B_{s'}.
\end{multline*}
An integration by parts shows that  the first line of the RHS of the formula is of order $\varepsilon$ and after plugging it into Eq.~(\ref{eq:6}) we obtain Eq.~(\ref{eq:7}).
\end{rem}

\section{Assumptions and basic results}
\label{basic}
We derive a solution of Eq.~(\ref{eq:1}) in the adiabatic limit $\varepsilon \to 0$ under three additional assumptions.
\begin{ass}
\begin{enumerate}
\item[(A)] For each $s$, $L_1(s) - 1/2 L_2^2(s)$ generates  a contraction semigroup and $i L_2(s)$ is self-adjoint,
\item[(B)] $L_1(s), s \in (0,\,1)$ is a family of operators for which $0$ remains a uniformly isolated discrete eigenvalue,
\item[(C)] $\ker L_2(s) \supseteq \ker L_1(s),\,s\in (0,1) $.
\end{enumerate}
\end{ass}

Condition (A) is a sufficient and necessary condition of a stochastic version of Hille-Yosida theorem; It implies that $U_\varepsilon(s,\,s')$ is a contraction, i.e. $||U_\varepsilon(s,\,s')|| \leq1$. This prevents an exponential blow up of solutions, and it is a standard in the adiabatic theory \cite{Joye, AFGG}. 

\begin{prop}[stochastic Hille-Yosida]
\label{shy}
Let $U(s,\,s')$ be a propagator associated to a stochastic differential equation
$$
\d U(s,s') = L_1 U(s,\,s') \d s + L_2 U(s,\,s') \d B_s, \quad U(s',s') = \id.
$$
Then the following are equivalent
\begin{enumerate}
\item[(i)] $U(s,\,s')$ is a contraction, i.e. $||U(s,\,s')|| \leq 1$,
\item[(ii)] $L_1 - 1/2 L_2^2$ is a generator of a contraction semigroup and $i L_2$ is self-adjoint.
\end{enumerate}
Furthermore (ii) implies that $L_1$ is a generator of contractions.
\end{prop}

{\bf Proof:} 
Without loss of generality we put $s' = 0$, and throughout the proof we denote $x(s) = U(s,\,0) x(0)$. The condition that $U(s,\,0)$ is a contraction is then equivalent to the statement that $||x(s)|| \leq ||x(0)||$ for all initial vectors $x(0)$. 

(ii) $\implies$ (i): By \^{I}to rules we have 
\begin{align*}
\d ||x(s)||^2 &= [ (L_1 x(s),\, x(s)) + (x(s),\, L_1 x(s)) + (L_2 x(s),\, L_2 x(s)) ] \d s \\
		   & \quad + [ (L_2 x(s), x(s)) + (x(s), L_2 x(s)) ] \d B_s \\
		  &= ((2 \mathrm{Re} L_1 - L_2^2)x(s), x(s)) \d s,
\end{align*}
where the last line is due to the assumption $L_2^* = - L_2$. Recall that $L_1 - 1/2 L_2^2$ is a generator of contraction on a Hilbert space if and only if it is dissipative, i.e. $\mathrm{Re}(L_1 - 1/2 L_2^2) \leq 0$. It then follows that $\d ||x(s)||^2 \leq 0$.

(i) $\implies$ (ii): We first prove that $L_2$ generates isometries, by proving that  both $L_2$ and $-L_2$ generate contraction semigroups. Suppose to the contrary that there exists an interval $I = (I_-,\,I_+)$ such that for $\varphi \in I$ and some $x \in \mathcal{H}$ we have $||e^{\varphi L_2} x|| > ||x||$. 

 We consider the same decomposition of $\d L$ as in (i). We treat $L_1 - 1/2 L_2^2$ as a perturbation and express $U(s) \equiv U(s,0)$ be a Duhamel formula. Since the perturbation is deterministic this is the standard version of the formula,
$$
U(s) = e^{L_2 B_s} + \int_0^s e^{L_2 (B_s - B_{s'})} (L_1 - \frac{1}{2} L_2^2) U(s') \d s'.
$$
The event $E_{I,\,s} = \{ B_s \in I \quad \mbox{and} \quad I_- - 1 \leq B_t \leq I_+ + 1,\,0\leq t \leq s\}$ has a non-zero probability for any interval $I$ and any $s$. By choosing $s$ sufficiently small we can then achieve $||U(s) x|| > ||x||$, which is in contradiction with (i).

Since $L_2^* = - L_2$ we have
$$
\d ||x(s)||^2= \left[ (L_1x(s),x(s)) + (x(s),L_1x(s)) + (L_2x(s), L_2x(s))\right] \d s
$$
and since $\d ||x(0)|| \leq 0$ we conclude that $L_1 - 1/2 L^2_2$ is dissipative and hence generate a contraction semigroup. 

The last claim of the proposition is not related to a classification of contraction semigroups. To prove it, observe that if $L_2$ is antiself-adjoint then $L_2^2$ is a generator of contractions. Hence $L_1 - 1/2 L_2^2 + 1/2 L^2_2$ is also a generator of contractions. Alternatively $L_1$ is a generator of the semigroup $\mathbb{E}[X_t]$.
\hfill $\square$

\begin{rem}
In a Banach space version of the proposition, the condition $i L_2$ is self-adjoint should be replaced by $L_2$ is a generator of isometries. The proof is technically more involved  and requires a version of Trotter-Kato formula that does not seem to be available in the literature (\cite{Kurtz} assumes compact state space, while \cite{Gough2, Durr} assume  the Hilbert space structure). In particular that if  $\d L_j,\,j=1,2$ is the generator of a propagator $U_j(s,s')$, then the propagator $U(s,s')$ generated by a sum $\d L_1 + \d L_2$ can be expressed as
$$
U(s,s') = \lim_{N \to \infty} U_1(s,s_N) U_2(s,s_N) U_1(s_N, s_{N-1}) U_2(s_N, s_{N-1}) \dots U_1(s_1,s') U_2(s_1,s'),
$$
where $s \geq s_N \geq  \cdots \geq s_1 \geq s'$ is any partition of the interval with a mesh going to $0$ as $N \to \infty$. This implies that if $\d L_1$ and $\d L_2$ generates contractions then so does $\d L_1 + \d L_2$.
\end{rem}

The gap condition, assumption (B), is also completely standard in the adiabatic theory. Since $L_1$ is a generator of contraction semigroup we have (see \cite{AFGG}) $\ker{L_1} \cap \ran{L_1} = 0$ and the gap condition implies
\begin{equation}
\label{decomposition}
\mathcal{B} = \ker L_1(s) \oplus \ran L_1(s).
\end{equation}

 The rather restrictive condition (C) allows to define the slow manifold and we cannot imagine how it can be relaxed.

 Before stating our results we shortly recall concepts from the adiabatic theory, see \cite{AFGG} or \cite{Te03} for a more thorough exposition. Let $P(s)$ be a $\mathcal{C}^1$ family of projections on $\mathcal{B}$ then the equation
 $$
 \pder{s} T(s,\,s') = [\dot P(s),\, P(s)] T(s,\,s'), \quad T(s',\,s') = \id
 $$
 defines parallel transport on $\ran P(s)$. The name ``parallel transport" is justified by two crucial properties
 \begin{enumerate}
 \item[(i)] $T(s,\,s') P(s') = P(s) T(s,\,s')$,\\
 \item[(ii)] A section $x(s) = T(s,\,s') x(s') \in \ran P(s)$ satisfies the equation
 $$
 P(s) \dot x(s) =0.
 $$ 
 \end{enumerate}
 
The parallel transport relevant to Eq.~(\ref{eq:1}) is given by the projection $P(s)$ on $\ker L_1(s) $ in the direction of $\ran L_1(s) $. This projection is well defined thanks to the decomposition Eq.~(\ref{decomposition}).
Henceforth $T(s,\,s')$ shall always refer to this particular projection, unless stated otherwise.

\begin{thm}
\label{thm:1}
Let $L_1(s),\,L_2(s)$ be $C^3$ families of operators satisfying assumptions (A)-(C). Then the differential equation $\varepsilon \,\d X(s) = \d L(s) X(s)$ admits solutions of the form 
$$
X(s) = a_0(s) + \sqrt{\varepsilon} \int_0^s U_\varepsilon(s,\,s') L_2(s') b_1(s')\, \d W_{s'} + \varepsilon (a_1(s) + b_1(s)) + O(\varepsilon^{\frac{3}{2}}),
$$
where
\begin{align*}
a_0(s) &= T(s,\,0) a_0(0),\\
b_1(s) &= L_1(s)^{-1} \dot{a_0}(s),\\
a_1(s) &= \int_0^s T(s,\,s') P(s') \dot{b}_1(s')\, \d s',
\end{align*}
and the initial condition $a_0(0)$ belongs to $\ker L_1(0)$. 
\end{thm}

We note that the integrand $U_\varepsilon(s,s')$ refers to the future and the integral is the backward \^{I}to integral. 
 The theorem is an immediate corollary of a more general Theorem~\ref{thwg} that describes the full expansion to all orders in $\varepsilon$. We feature it separately because we are not aware of any application of the expansion beyond the first order.
 
 \section{Stochastic Schr\"{o}dinger equation} 
 \label{schrodinger}

 The theorem may be applied to a driven stochastic Schr\"{o}dinger equation \cite[Chapter 5]{Holevo},
 $$
 \varepsilon \d \! \ket{\psi(s)} = -(i H(s) + \frac{1}{2} \Gamma(s)^2) \ket{\psi(s)} \, \d s - \sqrt{\varepsilon}i \Gamma(s) \ket{\psi(s)} \d B_s,
 $$
 where $\ket{\psi}$ is a vector in a Hilbert space and $H,\,\Gamma$ are self-adjoint operators. The equation generates unitary evolution and the average state $\bar \rho(s) = \mathbb{E}[\ket{\psi(s)}\!\bra{\psi(s)}]$ satisfies a Lindblad equation 
 \begin{equation}
 \label{eq:lindblad}
\varepsilon \dot{\bar{\rho}}(s) = -i[H(s),\bar\rho(s)] + \Gamma(s) \bar\rho(s) \Gamma(s) - \frac{1}{2}(\Gamma^2(s) \bar\rho(s) + \bar\rho(s) \Gamma^2(s)).
\end{equation}
 As in the deterministic case \cite[Section 3.1]{AFGG}, we need to subtract the dynamical phase before we can directly apply the adiabatic theorem. For an integrable function $E(s)$ and a square integrable function\footnote{The artificial square root in the definition of $\gamma$ was introduced in order to have the final results in the same form as in the Lindblad case.} $\sqrt{\gamma(s)}$ the transformation $ H(s) \to H(s) - E(s)$, $\Gamma(s) \to \Gamma(s) - \sqrt{\gamma(s)}$ transforms the solution of the stochastic Schr\"{o}dinger equation according to
 $$
 \ket{\psi(s)} \to e^{+i \frac{1}{\varepsilon} \int_0^s E(t) \d t + i \frac{1}{\sqrt{\varepsilon}} \int_0^s \sqrt{\gamma(t)} \d B_t} \ket{\psi(s)}.
 $$
 
  For simplicity we consider a $d$-dimensional Hilbert space and $H(s),\,\Gamma(s)$ with simple eigenvalues $E_0(s) = 0,\dots,E_{d-1}(s)$, $\sqrt{\gamma_0(s)} = 0,\,\dots,\,\sqrt{\gamma_{d-1}(s)}$ corresponding to a joint normalized eigenbasis $\ket{\psi_0(s)},\,\dots,\,\ket{\psi_{d-1}(s)}$. The eigenstate $\ket{\psi_k(s)}$ is determined only up to a phase and without loss of generality we assume that it is chosen in accordance with the parallel transport associated to the projection $\ket{\psi_k(s)}\! \bra{\psi_k(s)}$. Primarily, we shall study  solutions
 $\ket{\psi_\varepsilon(s)}$ of the stochastic Schr\"{o}dinger equation with an initial condition $\ket{\psi_\varepsilon(0)} = \ket{\psi_0(0)}$. Likewise we can study solutions with an initial condition $\ket{\psi_k(0)},\,k\in(1,\,d-1)$ after applying the above mentioned transformations. 
 
Of particular interest is the tunneling out of the ground state defined as
 \begin{align*}
 T_\varepsilon(s) &= 1 - |\braket{\psi_0(s)}{\psi_\varepsilon(s)}|^2 \\
 			  &= \sum_{k=1}^{d-1} |\braket{\psi_k(s)}{\psi_\varepsilon(s)}|^2.
 \end{align*}

  \begin{thm}
 Let $H(s),\,\Gamma(s)$ be as above. Then the stochastic Schr\"{o}dinger equation admits a solution  
 $$
 \ket{\psi_\varepsilon(s)} = \ket{\psi_0(s)} + \sqrt{\varepsilon} \sum_{k=1}^{d-1} \left(\int_0^s  D_\varepsilon^{(k)}(s,\,s')  t_k(s') \d B_{s'} \right) \ket{\psi_k(s)} + O(\varepsilon),  
 $$
 where
 $$
D_\varepsilon^{(k)}(s
,s')=e^{-i \frac{1}{\varepsilon} \int_{s'}^s E_k(t) \d t - i \frac{1}{\sqrt{\varepsilon}}\int_{s'}^s \sqrt{\gamma_k(t)} \d B_t} ,\quad t_k(s) = -i\sqrt{\gamma_k(s)}\frac{\braket{\psi_k(s)}{ \dot{\psi_0}(s)}}{-i E_k(s) - \frac{1}{2} \gamma_k(s)}. 
 $$
 In particular for the tunneling we have $T_\varepsilon(s) = \varepsilon \sum_{k=1}^{d-1}  T_k(s) + O(\varepsilon^\frac{3}{2})$,
 $$
 T_k(s) =  \left| \int_0^s D_\varepsilon^{(k)}(s,s') t_k(s') \d B_{s'} \right|^2 .
 $$
 In the leading order, terms $T_k(s)$ are independent random variables, and each term has an exponential distribution with mean 
 \begin{equation}
 \label{mean}
 \mathbb{E}[ T_k(s)] = \int_0^s |t_k(s')|^2 \d s'. 
 \end{equation}
 \end{thm}
 
 {\bf Proof:}
 Conditions  (A)-(C) for $L_1(s)  = -(i H(s) + \frac{1}{2} \Gamma(s)^2)$ and $L_2(s) = -i \Gamma(s)$ are clearly satisfied. $U_\varepsilon(s,s')$ is a unitary propagator. And the operator $L_1(s)$ has eigenvectors $\ket{\psi_k(s)}$ corresponding to simple discrete eigenvalues $-i E_k(s) - (1/2)\gamma_k(s)$.
    In view of Theorem~\ref{thm:1} and the discussion above we then have in the leading order
 \begin{equation}
 \label{p2.1}
 U_\varepsilon(s,\,s') \ket{\psi_k(s')} = D_\varepsilon^{(k)}(s,s') \ket{\psi_k(s)} + O(\sqrt\varepsilon).
 \end{equation}
 We proceed to the next order for the case with the initial condition $a_0(0) = \ket{\psi_0(s)}$. In order to do so we need to compute the coefficient $b_1(s)$. We express it in the joint eigenbasis of $H$ and $\Gamma$,
 $$
 b_1(s) = \sum_{k=1}^{d-1} \frac{\braket{\psi_k(s)}{\dot{\psi_0}(s)}}{-i E_k(s) - \frac{1}{2} \gamma_k(s)} \ket{\psi_k(s)}.
 $$
 It then follows from Theorem~\ref{thm:1} that 
 $$
 \ket{\psi_\varepsilon(s)} = \ket{\psi_0(s)} + \sqrt{\varepsilon} \sum_{k=1}^{d-1} \left(\int_0^s  U_\varepsilon(s,s')  t_k(s') \ket{\psi_k(s')} \d B_{s'} \right)  + O(\varepsilon),  
 $$
and by substituting from Eq.~(\ref{p2.1}) we obtain the first equation of the theorem. The expression for the tunneling is an immediate consequence. To compute the mean of the tunneling we use Formula~(\ref{itometry}).
It remains to show that transitions to different excited states are independent in the leading order and that the distribution of the tunneling is exponential. This will require some effort.
 
 We recall that exponential probability distribution with mean $\mu$ has a probability density function $p(x) = \mu^{-1} e^{-\frac{x}{\mu}}$ and is uniquely characterized by its moments $\int p(x) x^n = n! \mu^n$.  Our strategy is to compute the moments by establishing a recurrence relation between $\mathbb{E}[T_\varepsilon^n]$ and $\mathbb{E}[T_\varepsilon^{n-1}]$.
 
 For convenience we first express the tunneling as a forward stochastic integral. Using the computation in Remark~\ref{conversion}, Eq.~(\ref{eq:7}), with $L_1 = ( -i E_k - 1/2 \gamma_k)$ and $L_2 = -i \sqrt{\gamma_k}$ we have
 $$
 \int_0^s D^{(k)}_\varepsilon(s,\,s') t_k(s') \d B_{s'} = D_\varepsilon^{(k)}(s,0) \int_0^s D_\varepsilon^{(k)}(0,\,s') r_k(s') \d B_{s'} + O(\sqrt{\varepsilon}),
 $$
 where $r_k(s) = -i\sqrt{\gamma_k(s)} \braket{\psi_k(s)}{ \dot{\psi_0}(s)} /(-i E_k(s) + \frac{1}{2} \gamma_k(s))$. We hence obtain a forward expression for the tunneling in the leading order,
 $$
 T_k(s) =  \left| \int_0^s D_\varepsilon^{(k)}(0,s') r_k(s') \d B_{s'} \right|^2 .
 $$
 Note that $|t_k(s)|^2 = |r_k(s)|^2$, as it has to be for the mean to remain the same.
  
 We start by considering a single transition $T_k(s)$. \^{I}to rules imply
 $$
 \d T_k(s) = \left(\int_0^s \bar D_\varepsilon^{(k)}(0,s') \bar r_k(s') \d B_{s'}\right) D_\varepsilon^{(k)}(0,s) r_k(s) \d B_{s} + c.c. + |r_k(s)|^2 \d s,
 $$
 and
 $$
 (\d T_k(s))^2 =  \left(\int_0^s \bar D_\varepsilon^{(k)}(0,s') \bar r_k(s') \d B_{s'}\right)^2 D_\varepsilon^{(k)}(0,s)^2 r^2_k(s) \d s 
+c.c +  2 T_k(s) |r_k(s)|^2 \d s. 
 $$
 Using integral version of $\d T^n = n T^{n-1} \d T + (1/2)n(n-1) T^{n-2} \d T \d T$ and taking the expectation value we have (use the first formula in Eq.~(\ref{itometry}))
 \begin{multline*}
 \mathbb{E}[T_k^n(s)] = n \int_0^s \mathbb{E}[T_k^{n-1}(s')] |r_k(s')|^2 \d s' + n (n-1) \int_0^s \mathbb{E}[T_k^{n-1}(s')] |r_k(s')|^2 \d s' \\
 + \frac{n(n-1)}{2} \int_0^s \mathbb{E}[T^{n-2}(s') \left(\int_0^{s'} \bar D_\varepsilon^{(k)}(0,s'') r_k(s'') \d B_{s''}\right)^2 D_\varepsilon^{(k)}(0,s')^2 r^2_k(s') \d s'] + c.c.
 \end{multline*}
 Integrating by parts with respect to the factor $e^{ -i \frac{2}{\varepsilon} \int_{0}^{s'} E_k(t) \d t} $ shows that the second line
 is of order $\varepsilon^{1/2}$, whence
 $$
 \mathbb{E}[T^n_k(s)] = n^2 \int_0^s \mathbb{E}[T_k^{n-1}(s')] |r_k(s')|^2 \d s' + O(\varepsilon^{\frac{1}{2}}).
 $$
 Using this relation recursively we arrive at
 \begin{align*}
 \mathbb{E}[T^n_k(s)]    &= (n!)^2 \int\limits_{0\leq s_1 \leq \dots \leq s_n \leq s} \prod_{i=1}^n |r_k(s_i)|^2 \d s_1 \dots \d s_n + O(\varepsilon^\frac{1}{2}) \\
 				      &= n! \left(\int_0^s |r_k(s')|^2 \d s'\right)^n + O(\varepsilon^\frac{1}{2}),
 \end{align*}
 which is exactly the relation characterizing exponential distribution.
 
 Now consider two terms $T_k(s),\,T_l(s)$ for $l \neq k$. By \^{I}to formula we have
 \begin{align*}
 \mathbb{E}[T_k(s) T_l(s)] &= \int_0^s \left(\mathbb{E}[\d T_k(s') T_l(s')] + \mathbb{E}[T_k(s') \d T_l(s')] + \mathbb{E}[\d T_k(s') \d T_l(s')] \right) \\
 					&=\int_0^s \left(|r_k(s')|^2\mathbb{E}[T_l(s')] + |r_l(s')|^2\mathbb{E}[T_k(s')] \right) \d s'+ O(\sqrt{\varepsilon}) \\
					&= \mathbb{E}[T_k(s)] \mathbb{E}[T_l(s)] + O(\sqrt{\varepsilon}).
 \end{align*}
 That the last term on the RHS of the first line is of order $\varepsilon^{1/2}$ can be shown by integration by parts. Hence we showed that $T_k$ and $T_l$ are uncorrelated and we proceed to higher powers by induction. Suppose that $T_k^{n-1}$ and $T_l^m$ ($T_k^n$ and $T_l^{m-1}$)  are uncorrelated to the leading order, then we have 
 \begin{align*}
 \mathbb{E}[T_k^n T_l^m] &= \int \left( \mathbb{E}[\d (T_k^n) T_l^m]  + \mathbb{E}[T_k^n \d(T_l^m)] + \mathbb{E}[\d (T_k^n) \d (T_l^m)] \right) \\
 						&= \int \left( n^2 |r_k|^2 \mathbb{E}[T_k^{n-1} T_l^m]  + m^2 |r_l|^2 \mathbb{E}[T_k^{n} T_l^{m-1}] \right) + O(\sqrt{\varepsilon}) \\
						&= \int \left(\d (\mathbb{E}[T_k^n])\mathbb{E}[T_l^n] + \mathbb{E}[T_k^n] \d (\mathbb{E}[T_l^n]) \right) + O(\sqrt{\varepsilon} ) \\
						&=\mathbb{E}[T_k^n]\mathbb{E}[T_l^n] + O(\sqrt{\varepsilon}).
 \end{align*}
 So to leading order $T_k$ and $T_l$ are independent, which finishes the proof.
  \hfill $\square$
 
 \begin{rem}
The main deficiency of the expansion in Theorem~\ref{thm:1} is that it involves the propagator itself, albeit in a higher order.   It is straightforward, although cumbersome, to recursively eliminate the propagator. 
 We do not know of any more direct manner to derive  higher order terms in the expansion. 
 \end{rem}
 
 Formula (\ref{mean}) for the mean tunneling has been derived in \cite{AFGG} using the corresponding adiabatic Lindblad equation, Eq.~(\ref{eq:lindblad}),  and subsequently used to study an optimal sweeping rate \cite{AFGG2} and Landau-Zener tunneling with dephasing \cite{AFGG3}. The mean tunneling is additive, which was interpreted as the tunneling in the dephasing case being local and unidirectional. The full statistics of the tunneling derived here, offers an unexpected twist. If the tunneling was additive it would have a Gaussian distribution, not an exponential one. It follows that only the mean tunneling is additive, while higher order cumulants exhibit non-local behavior typical for the Hamiltonian evolution. 
 
 \section{Full expansion and its proof}
 \label{main}
 
 Now we present the main theorem, that describes the expansion to all orders.
 
 \begin{thm} \label{thwg}
Let $L_1(s), L_2(s)$ be $C^{N+2}$-families of operators satisfying Assumptions~(A)-(C). Then
 \begin{enumerate} 
  \item The differential equation $\varepsilon\, \d X=\d L(s)X$ admits solutions of the form 
\begin{align} \label{solu}
 X(s)=\sum_{n=0}^N\varepsilon^n\left( \varepsilon^{-1/2}\int_0^s U_\varepsilon(s,\,s') L_2(s') b_n(s')\, \d B_{s'} + a_n(s)+b_n(s)\right)+\varepsilon^{N}r_N(\varepsilon,s)
\end{align}
with
\begin{itemize}
 \item $a_n(s) \in \ker L_1(s), \ b_n(s) \in \ran L_1(s)$.
\item initial data $x(0)$ is specified by arbitrary $a_n(0) \in \ker L_1(0)$; however, the $b_n(0)$ are determined
below by the $a_n(0)$ and together define the "slow manifold".
\end{itemize}
\item The coefficients are determined recursively through $(n=0, \dots, N)$
\begin{align}
 b_0(s)&=0\,, \nonumber \\
a_n(s)&=T(s,0)a_n(0)+ \int_0^sT(s,s')\dot{P}(s')b_n(s')ds'\,, \label{an} \\
b_{n+1}(s)
&=L_1(s)^{-1}\left(\dot{P}(s)a_n(s)+P_{\perp}(s)\dot{b}_n(s)\right)\,. \label{bnpo}
\end{align}
\item The remainder is uniformly small in $\varepsilon$ and is of the form
$$
r_N(\varepsilon,s) = \sqrt{\varepsilon} \int_0^s r_N^{(2)}(\varepsilon, s') \d B_{s'} + \varepsilon r_N^{(1)}(\varepsilon,s) ,
$$
where $r_N^{(1)}(\varepsilon, s),\,r_N^{(2)}(\varepsilon, s)$ are uniformly bounded  functions. In particular, $r_N(\varepsilon,s) = O(\sqrt{\varepsilon})$.
\end{enumerate}
\end{thm}

{\bf Proof:}
Since $L_1(s)$ is a generator of a contraction semigroup (see the last claim in Proposition~\ref{shy}) we can use the standard deterministic adiabatic theory for an equation $\varepsilon \d \tilde{X}(s) = L_1(s) \tilde{X}(s)$. Using the expansion in \cite[Theorem~6]{AFGG} the equation has a solution,
$$
\tilde{X}(s) = \sum_{n=0}^N\varepsilon^n\left( a_n(s)+b_n(s)\right)+\varepsilon^{N+1}r^{(1)}_N(\varepsilon,s),
$$
where $r^{(1)}_{N}(\varepsilon,s)$ is uniformly bounded.

By the Duhamel formula of Lemma~\ref{duhamel} we then have a solution of the stochastic equation,
\begin{multline*}
 X(s)=\sum_{n=0}^N\varepsilon^n\left( \varepsilon^{-1/2}\int_0^s U_\varepsilon(s,\,s') L_2(s') b_n(s')\, \d B_{s'} + a_n(s)+b_n(s)\right) \\+\varepsilon^{N+1}r^{(1)}_N(\varepsilon,s) + \varepsilon^{N+\frac{1}{2}} \int_0^s U_\varepsilon(s,\,s') L_2(s') r^{(1)}_N(\varepsilon,s') \d B_{s'}.
\end{multline*}
This is exactly the expansion of the theorem with $r_N^{(2)}(\varepsilon, s') = U_\varepsilon(s,\,s') L_2(s') r^{(1)}_N(\varepsilon,s')$. 

That $r_n^{(2)}(\varepsilon,s)$ is uniformly bounded (with probability $1$) follows from the assumption (A), which implies that $||U_\varepsilon(s,s')|| \leq 1$. That the error is of the order $O(\sqrt\varepsilon)$ follows from Lemma~\ref{bound}, or more precisely from a backward integration counterpart of the lemma. In fact, Lemma~\ref{bound}.(a) is sufficient for that conclusion, at the same time Lemma~\ref{bound}.(b) gives better error estimates.
\hfill $\square$

We conclude with several remarks regarding the generality of our exposition. Including several independent noises, i.e. $L_2 \d B \to \sum_k L_2^{(k)} \d B_k$ where $B_k$ are independent Brownian motions, is straightforward. In particular the tunneling Eq.~(3) turns into a sum over the noises, each giving an independent contribution to the tunneling. Boundedness of $L_1,\,L_2$ can surely be relaxed as well as the gap condition, Assumption~(B). We do not plan to elaborate on any of these generalizations. On the other hand it is important to allow generators $L_1(s),\,L_2(s)$ to depend on the Brownian motion, $B_t$, for $s \geq t \geq 0$. We hope to address this question in a further work. 
 
\medskip\noindent
{\bf Acknowledgements.} 
I thank Gian Michele Graf and Eddy Mayer-Wolf for  fruitful discussions. A part of the work was done while I visited the Isaac Newton Institute in Cambridge, UK. Support by the Swiss National Science Foundation is acknowledged.

\bibliography{draft1.bib}
\bibliographystyle{plain.bst}


\end{document}